# Electrowetting on a semiconductor

Steve Arscott[a] and Matthieu Gaudet

*Institut d'Electronique, de Microélectronique et de Nanotechnologie (IEMN), CNRS UMR8520, The University of Lille, Cité Scientifique, Avenue Poincaré, 59652 Villeneuve d'Ascq, France.*

[a]steve.arscott@iemn.univ-lille1.fr

We report electrowetting on a semiconductor using of a mercury droplet resting on a silicon surface. The effect is demonstrated using commercial n-type and p-type single-crystal (100) silicon wafers of different doping levels. The electrowetting is reversible - the voltage-dependent wetting contact angle variation of the mercury droplet is observed to depend on both the underlying semiconductor doping density and type. The electrowetting behaviour is explained by the voltage-dependent modulation of the space-charge capacitance at the metal-semiconductor junction – current-voltage and capacitance-voltage-frequency measurements indicate this to be the case. A model combining the metal-semiconductor junction capacitance and the Young-Lippmann electrowetting equation agrees well with the observations.

Two effects in science - which up until today have been very separate – are rectification[1] discovered by Ferdinand Braun in 1874 and electrowetting[2] observed by Gabriel Lippmann in 1875. Solid-state rectification - using a *metal-semiconductor* contact,[3] i.e. a diode - is one of the workhorses of the microelectronics revolution.[4] Whilst modern electrowetting-on-dielectric[5] – using a *liquid-insulator-conductor* junction – has many applications[6-14] including displays,[6] droplet transport,[7,8] smart optics,[9,10] electronic paper,[11,12] miniaturized chemistry[13] and energy harvesting.[14] Here, electrowetting on a semiconductor is reported which relies on the voltage-dependent capacitance of a space-charge zone created at a reverse biased metal-semiconductor junction.[15,16] The effect is demonstrated using a mercury droplet resting on a commercial, single-crystal silicon wafer. Current-voltage and capacitance-voltage measurements indicate the importance of the space-charge region in the effect - a model combining the metal-semiconductor junction capacitance[4] and the Young-Lippmann electrowetting equation[5] agrees well with the observations.

Electrowetting[5] can be understood using the Young-Lippmann equation:

$$\cos\theta = \cos\theta_0 + \frac{1}{2\gamma}C_a V^2 \qquad (1)$$

where $\theta$ is the contact angle of a droplet resting on a surface – often composed of an evaporated metal coated with a thin hydrophobic insulating layer, $\theta_0$ is the contact angle of the droplet at zero bias, $\gamma$ is the surface tension of the liquid, $C_a$ is the *areal* capacitance of the insulating layer (in Farads per square meter) and $V$ is the applied voltage. In the case of electrowetting-on-dielectric (EWOD), $C_a$ does not vary with voltage. However, in the case of a liquid-insulator-*semiconductor* system, it has been shown that the value of $C_a$ – and thus $\theta$ - depends strongly on the applied voltage magnitude *and* polarity - related to the specific doping density and type of the semiconductor.[17] Under reverse bias, it has also been shown that $C_a$ – and thus $\theta$ - also depends on illumination.[17] For liquid-semiconductor interfaces, the value of $C_a$ is also dependent on the magnitude of the applied voltage and its polarity - depending on the doping level and type of the underlying semiconductor. Such interfaces have been successfully employed for commercial semiconductor profiling systems.[18]

Fig. 1 illustrates the band structures of a junction – referred to as a *Schottky* contact[4] - formed by placing a drop of conducting liquid onto the surface of a semiconductor. An n-type semiconductor in thermal equilibrium [Fig. 1(a)] – zero bias, and under reverse bias [Fig.

1(b)] – a negative voltage applied to the droplet. A p-type semiconductor in thermal equilibrium [Fig. 1(c)] – zero bias, and under reverse bias [Fig. 1(d)] – a positive voltage applied to the droplet. At thermal equilibrium the Fermi level (short dashed line) in the semiconductor lines up with the top of the conduction band in the conducting droplet (shaded region) - the droplet will form a zero-bias contact angle $\theta_0$ with the surface of the semiconductor. Under reverse bias, the capacitance of the space-charge region decreases causing the contact angle of the conducting droplet – indicated by the long dashed line – to decrease as the reverse bias is increased.

It is well known that near-ideal Schottky diodes can be formed using *solid metal*-semiconductor junctions.[19,20] However, in order to demonstrate electrowetting at a Schottky junction a *liquid metal*-semiconductor contact is necessary - this system can be realized by forming an Hg-Si junction.[21-27]

Commercial, 3-inch diameter - 380 µm thick - (100) silicon wafers (Siltronix, France) were used for the experiments having two doping types: p-type (resistivity $\rho$ = 5-10 Ω cm) and n-type ($\rho$ = 5-10 Ω cm). The silicon wafers were cleaned and deoxidized using $H_2SO_4/H_2O_2$ and HF based solutions. For p-type silicon, removal of the native oxide prior to mercury droplet deposition is known to produce a good Schottky contact via a reduction of surface charges.[25] It should also be noted that removal of the oxide prior to deposition of the Hg contact produces a metal-semiconductor[22,23,25,27] contact whereas forming a thin oxide[21-23,26] prior to deposition of the mercury produces – strictly speaking – a metal-oxide-semiconductor (MOS) junction. Ohmic contacts were formed on the rear surface of the silicon wafers using ion implantation: Boron -$10^{20}$ cm$^{-3}$ – for the p-type wafers, and Phosphorous – $10^{20}$ cm$^{-3}$ – for the n-type wafers). All fabrication was performed in a class ISO 5/7 cleanroom (*T* = 20°C±0.5°C; *RH* = 45%±2%)**.**

Small droplets of mercury – having a diameter less than the capillary length ~1.9 mm - were placed directly onto the silicon surfaces. A DC voltage was applied to the mercury droplet via a metal probe and using an E3634A Power Supply (Agilent, USA). The reverse bias voltage was ramped (5 V s$^{-1}$) from 0 V to -50 V (n-type) and 50 V (p-type). The contact angle of the mercury droplet was recorded as a function of applied voltage. The contact angle data was gathered using a commercial Contact Angle Meter (GBX Scientific Instruments, France) - the contact angles were extracted using an interpolation model.[28]

Fig. 2 shows photographic evidence of electrowetting on a semiconductor using a mercury droplet on a silicon surface. Fig. 3 shows the variation of droplet contact angle with applied voltage (open circles data) for n-type and p-type silicon. The expected electrowetting effect was observed for all wafers tested. The contact angle reduces upon increasing reverse bias – by ~2.5° for n-type (0 to -20 V) and ~5° for p-type (0 to 30 V). The electrowetting occurs only under reverse bias - negative voltages applied to the mercury droplet for an n-type wafer and positive voltages applied to the mercury droplet for a p-type wafer. The role of the semiconductor doping density $N$ is also apparent from the results; the higher doped p-type wafer has a larger contact angle variation. The droplet contact radius $r$ is seen to increase when increasing the reverse bias voltage - the value of droplet contact radius variation is greater for the higher doped p-type wafer. The droplet contact surface is observed to increases significantly upon application of reverse bias voltage: ~10% for the n-type wafer (at -20 V) and ~15% for the p-type wafer (at 40 V); this observation could have implications for mercury based contact measurements in semiconductor studies – see later. The electrowetting is reversible – turning off the reverse bias enables the zero bias contact angles to be restored. Contact angle saturation[29,30] is also apparent from the data - for the n-type silicon, saturation begins at -20 V whereas for the p-type silicon it begins at 30 V. Contact angle saturation for EWOD is a well a documented phenomenon[29] which still defies a total understanding.[30] The measured contact angle variation indicates that the role of the voltage-dependent junction capacitance is important. However, to prove that the electrowetting behaviour is related to the reversed bias Schottky junction we have conducted electrical characterization.

Current-voltage (I-V) measurements - see Fig. 4 - were performed on the Hg-Si Schottky junctions used for the electrowetting experiments. The purpose of the I-V measurements was to be sure that the Hg-Si systems - used for the electrowetting experiments - behave as Schottky diodes in reverse bias.[21-27] The I-V measurements were performed using a 2612 System SourceMeter® (Keithley, USA). The *zero-bias* junction surfaces were measured to be $1\times10^{-6}$ m$^2$ (n-type) and $6.7\times10^{-7}$ m$^2$ (p-type).

The I-V measurements indicate that the Hg-Si contacts are functioning as reverse bias Schottky diodes. The measured current densities $J$ – plotted in Fig. 4 - are consistent with previous studies using silicon wafers having similar restivities.[21-27] For the Hg-nSi contact, $J$ = $2.3\times10^{-5}$ A m$^{-2}$ at -1 V rising to 11 A m$^{-2}$ at -15 V and 50 A m$^{-2}$ at -20 V  [Fig. 4(a)] – comparable to the litterature[21,24,25] following surface treatment with HF. For the Hg-pSi contact, we observe $J = 2\times10^{-3}$ A m$^{-2}$ at 1 V rising to 3.7 at 15 V and 22.4 A m$^{-2}$ at 25 V [Fig.

4(b)] – comparable to the litterature[21-23] following surface treatment with HF. The apparent reverse bias breakdown voltages are comparable with those given in ref. [22].

The I-V measurements also suggest an explanation for the observed contact angle hysteresis – see Fig. 3 (open circle data). The higher reverse bias leakage currents at voltages above -20 V - for the n-type - and 25 V - for the p-type - suggest that the space-charge zone can no longer be considered to be a near perfect capacitor – in this case, Eq. (1) no longer holds and one would not expect the contact angle to decrease when increasing the reverse bias voltage further.

Small signal capacitance-voltage-frequency (C-V-f) measurements – shown in Fig. 5 - were performed using the silicon wafers used for the electrowetting experiments for two reasons: (i) to evaluate the precise *areal* capacitance as a function of applied voltage and frequency and (ii) to measure the doping density level of the silicon wafers. Depending on pre-treatment of the silicon surface,[3] aluminium is known to form good Schottky contacts to both n and p-type silicon.[20] Use of a *solid* metal contact allows the extraction of $C_a$ as the surface area is constant – unlike Hg, enabling comparison with the contact angle data.

Small-signal C-V-f measurements were carried out using a Precision Impedance Analyzer 4294A (Agilent, USA) using a bias voltage of ±40V. A full calibration (open circuit – load (200Ω) – short circuit) was performed using a P/N101-190 S/N33994 Impedance Standard Substrate (Cascade Microtech, USA) over the frequency range (500 Hz – 1MHz) prior to the measurements. The high frequency C-V-f measurements allow the extraction[4] of the doping density of the wafers via a Mott-Schottky $1/C^2$ versus $V$ plot shown in the insets to Fig. 5 - at 1 MHz the measured capacitance is equal to the depletion layer capacitance as charge at the interface cannot follow the ac signal.[4] The doping levels of the n-type and p-type wafers were evaluated to be $5.22 \times 10^{14}$ cm$^{-3}$ (8.66 Ω cm) (max: $5.47 \times 10^{14}$ cm$^{-3}$, min: $5 \times 10^{14}$ cm$^{-3}$) and $2.12 \times 10^{15}$ cm$^{-3}$ (6.46 Ω cm) (max: $2.18 \times 10^{15}$ cm$^{-3}$, min: $2.05 \times 10^{15}$ cm$^{-3}$).

The C-V-f measurements also allow us to evaluate the series resistance $R_s$ introduced by the finite conductivity of the silicon wafers. We measure $R_s$ = 850 Ω (n-type) and 425 Ω (p-type) for Al-Si diodes having a surface area are comparable to the junction area of the Hg-Si diodes used for the electrowetting experiments.

By injecting the measured small signal areal capacitance values - obtained from the C-V-f measurements – into the Young-Lippmann equation given in Eq. (1), one obtains the thick dashed lines indicated on Fig. 3. This data is seen to fit well with the contact angle variation obtained from the electrowetting measurements (open circles) using the low-frequency (100 kHz) areal capacitance measurements indicating that the electrowetting is determined by the space-charge region of the Schottky diode. It is well documented that the capacitance of a space-charge region is dependent on measurement frequency[31] and depends strongly on interface traps.[32] Thus, the observations here suggest that the areal capacitance that the droplet '*sees*' during the electrowetting measurements is the *low-frequency areal capacitance* measured using the C-V-f measurements. On the other hand, the doping density extraction is achieved using a high frequency measurement as the traps do not have time to influence the capacitance of the depletion zone.[31,32]

A simple model for electrowetting at a Schottky junction can be formulated by employing an analytical model for the junction capacitance of a metal-semiconductor contact:[33]

$$C_a = \sqrt{\frac{\varepsilon_r \varepsilon_0 q N}{2(V_{bi} - V)}} + q D_{it} \qquad (2)$$

where, $\varepsilon_r$ is the dielectric constant of the semiconductor, $\varepsilon_0$ is the permittivity of free space, $q$ is the elementary charge, $N$ is the doping density in the semiconductor, $V_{bi}$ is built-in potential, $V$ is the applied voltage and $D_{it}$ is the interface trap density. Eq. (2) makes the assumption that $|V|$ and $V_{bi} \gg kT/q$ for electrowetting at 300K. Also, if the series resistance $R_s$ is non-negligible – due to the finite resistivity of the silicon wafer - $C_a$ is divided by a factor $\alpha = 1+qR_sI_R/kT$ where $I_R$ is the reverse bias leakage current, $T$ is the temperature and $k$ is Boltzmann's constant.[33]

Combining the Eq. (1) with Eq. (2) gives a simple model for electrowetting at a Schottky junction:

$$\cos\theta = \cos\theta_0 + \frac{1}{\alpha}\left[\sqrt{\frac{\varepsilon_r \varepsilon_0 q N}{8\gamma^2(V_{bi}-V)}} + \frac{qD_{it}}{2\gamma}\right]V^2 \qquad (3)$$

Note that $\cos\theta$ is not proportional to $V^2$ - as is the case for EWOD,[5] and that the electrowetting depends on the properties of the underlying semiconductor (doping type, $N$ and $\varepsilon_r$), the quality of the interface ($D_{it}$) and the leakage current $I_R$ flowing through the associated series resistance $R_s$. Eq. (3) holds only in reverse bias; under forward bias, conduction occurs and we can consider $C_a$ to be effectively zero and the droplet will not spread out – as is seen in the measurements (Fig. 3).

By using the measured values of the doping density $N$ and the series resistance $R_s$ – obtained from the C-V measurements, the measured leakage currents $I_R$ – obtained from the I-V measurements, a surface tension of mercury equal to 486.5 mJ m$^{-2}$, setting $V_{bi}$ to 0.7 V,[21-27] and by fitting the value of $D_{it}$ (n-type - $4\times10^{14}$ m$^{-2}$ V$^{-1}$; p-type – $2.4\times10^{14}$ m$^{-2}$ V$^{-1}$), Eq. (3) is plotted as a thin solid curves shown in Fig. 3. The fitted values for $D_{it}$ agree with measured values for the Hg-Si system.[25-27] Although contact angle saturation[29,30] is not considered here, Eq. (3) predicts contact angle saturation due to the increase in $I_R$ flowing through $R_s$ as $V$ is increased. Note also that if $\alpha = 1$ and $D_{it} = 0$, Eq. (3) describes electrowetting at an *ideal* Schottky junction.

We now know that for the Hg-Si junction a voltage-induced modification of the contact angle - at constant droplet volume $v_d$ - results in a modification of the surface area of the liquid-solid interface $A_{ls}$. As a consequence the measured current density $J = I A_{ls}$ – obtained from the I-V measurements - and the capacitance $C = C_a A_{ls}$ – obtained from the C-V measurements - should be modified using the following equation:

$$A_{ls} = \left[\frac{2\sqrt{\pi} v_d \sin^3\theta}{(2+\cos\theta)(1-\cos\theta)^2}\right]^{2/3} \qquad (4)$$

This effect should be taken into consideration for electrical measurements using conducting liquid-semiconductor (e.g. Hg-Si)[25] and conducting liquid-insulator-semiconductor (e.g. Hg-SiO$_2$-Si)[25] systems if surface restriction techniques – such as an o-ring[18,22] - are not employed to fix the diode junction area. In order to obtain the true values of $J$ (A m$^{-2}$) and $C_a$ (F m$^{-2}$) from the measured $I$ (A) and $C$ (F), one should divide the measured values by the voltage-dependent contact area $A_{ls}$ - obtained via electrowetting experiments - and not simply by the $A_{ls}$ at *zero bias*. By using Eq. (3) and Eq. (4) one can calculate the real current densities from the I-V measurements on the Hg-Si systems – shown as the modified thin curves in Fig. 4.

1. F. Braun, Ann. Phys. Chem. **153**, 556 (1874).

2. G. Lippmann, Ann. Chim. Phys. **5**, 494 (1875).

3. R. T. Tung, Mat. Sci. Eng. R **35**, 1 (2001).

4. S. M. Sze, *Physics of semiconductor devices* - 3rd Edition (Wiley-Interscience, New Jersey, 2007).

5. F. Mugele and J.-C. Baret, J. Phys. Condensed Matter. **17**, R705 (2005).

6. G. Beni and S. Hackwood, Appl. Phys. Lett. **38**, 207 (1981).

7. M. G. Pollack, R. B. Fair, and A. D. Shenderov, Appl. Phys. Lett. **77**, 1725 (2000).

8. J. Lee and C. -J. Kim, J. Microelectromech. S. **9**, 171 (2000).

9. B. Berge and J. Peseux, Eur. Phys. J. E **3**, 159 (2000).

10. T. Krupenkin, S. Yang, and P. Mach, Appl. Phys. Lett. **82**, 316 (2003).

11. R. A. Hayes and B. J. Feenstra, Nature **425**, 383 (2003).

12. H. You and A. J. Steckl, Appl. Phys. Lett. **97**, 023514 (2010).

13. A. R. Wheeler, Science **322**, 539 (2008).

14. T. Krupenkin and J. A. Taylor, Nat. Commun. **2**, 448 (2011).

15. N. F. Mott, Proc. Cambr. Philos. Soc. **34**, 568 (1938).

16. W Schottky, W. Naturwissenschaften **26**, 843 (1938).

17. S. Arscott, Sci. Rep. **1**, 184 (2011).

18. P. Blood, Semicond. Sci. Technol. **1**, 7 (1986).

19. M. P. Lepselter and S. M. Sze, Bell System Tech J. **47**, 195 (1968).

20. H. C. Card, IEEE Trans. Electron. Dev. **23**, 538 (1976).

21. D. K. Donald, J. Appl. Phys. **34**, 1758 (1963).

22. P. J. Severin and G. J. Poodt, J. Electrochem. Soc. **119**, 1384 (1972).

23. M. Wittmer and J. L. Freeouf, Phys. Rev. Lett. **69**, 2701 (1992).

24. Q. Wang, D. Liu, J. T. Virgo, J. Yeh, and R. J. Hillard, J. Vac. Sci. Technol. A **18**, 1308 (2000).

25. H. J. Hovel, Solid-State Electronics **47**, 1311 (2003).

26. Y-J. Liu and H-Z. Yu, J. Phys. Chem. B, **107**, 7803 (2003).

27. J. Y. Choi, S. Ahmed, T. Dimitrova, J. T. C. Chen, and D. K. Schroder, IEEE Trans. Electron. Dev. **51**, 1164 (2004).

28. A.F. Stalder, G. Kulik, D. Sage, L. Barbieri, and P. Hoffmann Colloids Surf. A **286**, 92 (2006).

29. F. Mugele, Soft Matter **5**, 3377 (2009).

30. S. Chevalliot, S. Kuiper and J. Heikenfeld, J. Adhesion Sci. Technol. 1-26 (2011) DOI:10.1163/156856111X599580

31. E. H. Nicollian and A. Goetzberger, Bell System Tech. J. **46**, 1055 (1967).

32. J. Werner, K. Ploog and H. J. Queisser, Phys. Rev. Letts. **57**, 1080 (1987).

33. S. Pandey and S. Kal, Solid State Electron. 42, 943 (1998).

Figure 1. Energy band diagrams of electrowetting of a conducting droplet (grey) at a Schottky junction under different bias conditions. (a) n-type at thermal equilibrium, (b) n-type under reverse bias –a *negative voltage* applied to the droplet, (c) p-type at thermal equilibrium and (d) p-type under reverse bias – a *positive voltage* applied to the droplet. $E_c$ – bottom of the conduction band, $E_F$ – Fermi level, $E_v$ – top of the valence band and $V_{bi}$ = built-in voltage of the junction.

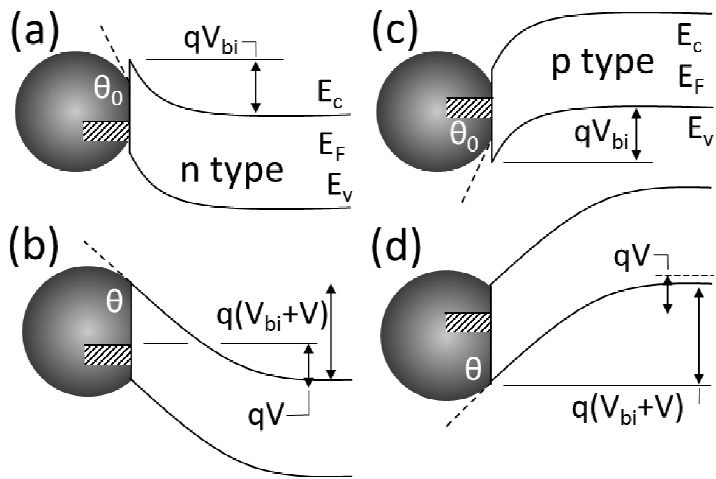

Figure 2. Evidence of electrowetting at the mercury-silicon Schottky junction. (a) n-type silicon at zero bias, (b) n-type silicon at a reverse bias of -20 V, (c) p-type silicon at zero bias and (d) p-type silicon at a reverse bias of +40 V. The measured contact angles $\theta$ (deg) and radii $r$ (μm) of droplet contact surface are given.

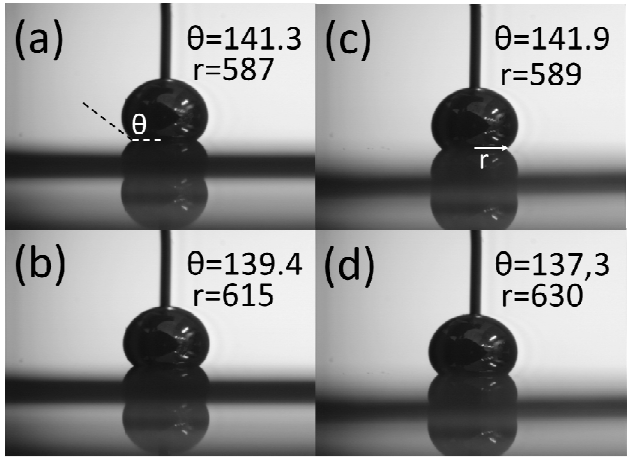

Figure 3. Contact angle versus voltage for the mercury-silicon Schottky junctions. (a) n-type silicon and (b) p-type silicon (open circles). Thick dashed lines obtained by injecting the measured areal capacitances obtained from the C-V-f measurements into the Young-Lippmann - Eq. (1). Thin solid lines indicate the modelled curves based our model - Eq. (3).

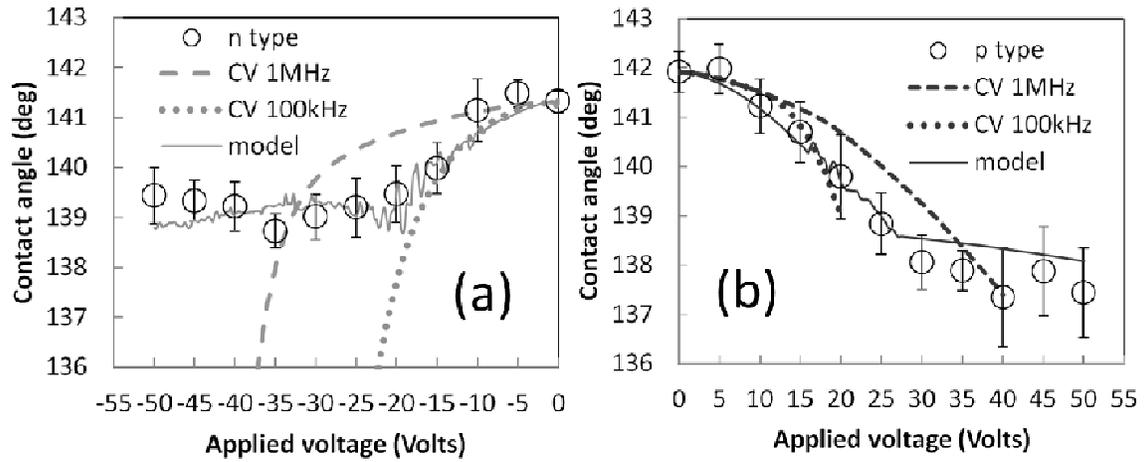

Figure 4. Measured current-voltage characteristics for mercury-silicon Schottky diodes. (a) n-type silicon and (b) p-type silicon. The thin lines indicate the true current density taking into account the voltage-dependent surface area given by our model – Eq. (3) and Eq. (4).

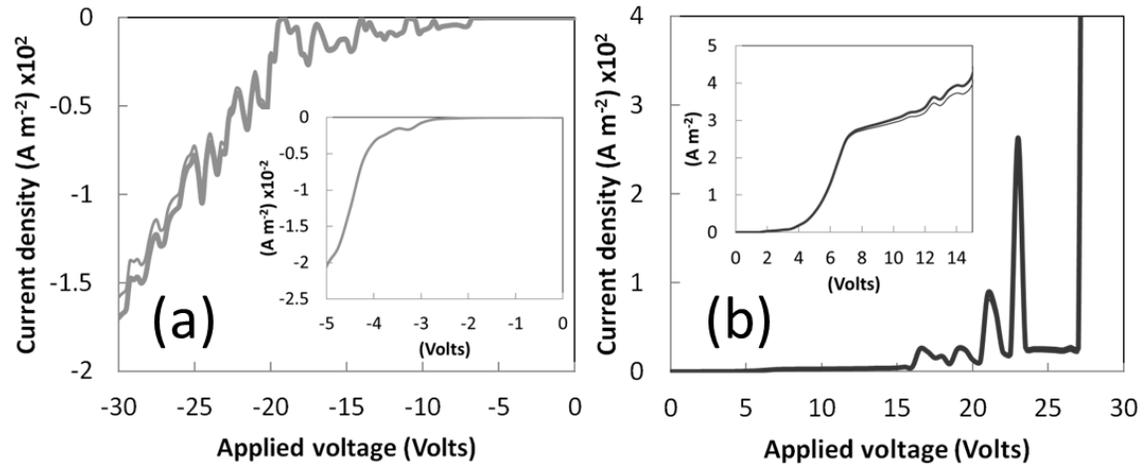

Figure 5. Measured capacitance-voltage-frequency curves for aluminium-silicon Schottky diodes formed using the same silicon wafers used for the electrowetting experiments. (a) n-type silicon and (b) p-type silicon. Measurement frequency = 100 kHz and 1 MHz. Insets show Mott-Schottky $1/C^2$ vs. $V$ plots at 1 MHz. The dashed curves in the insets are based on Eq. (2) with $D_{it} = 0$.

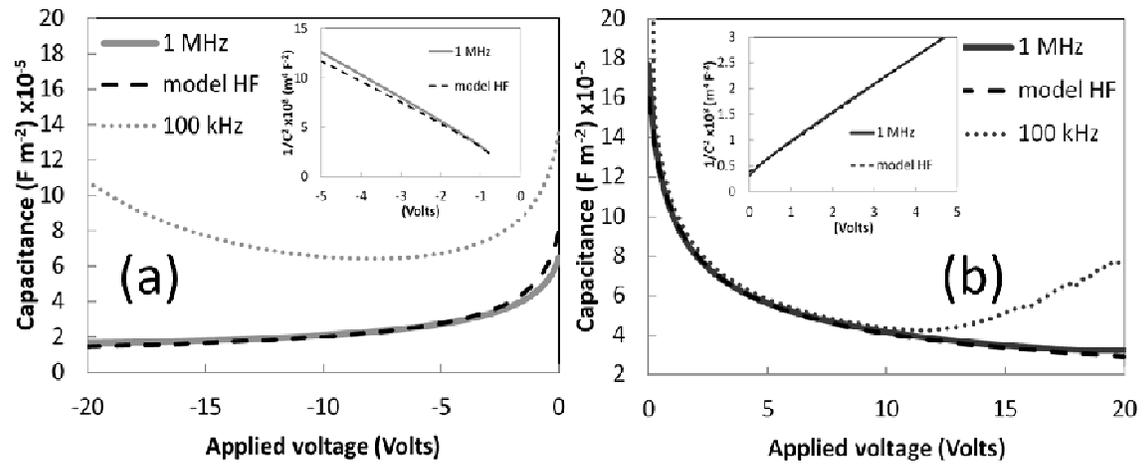